\documentclass[superscriptaddress,showpacs,twocolumn,prl,floatfix]{revtex4-1}
\usepackage[english]{babel}
\usepackage{units}
\usepackage{mathrsfs}
\usepackage{amsmath}
\usepackage{amssymb}
\usepackage{graphicx}
\usepackage{esint}
\usepackage{ulem}
\usepackage{mathrsfs}
\usepackage{dcolumn}
\usepackage{bm}
\usepackage{color}
\def\be{\begin{equation}}
\def\ee{\end{equation}}
\def\ber{\begin{eqnarray}}
\def\eer{\end{eqnarray}}

\def\kv{{\bf k}}

\def\Mv{{\bf M}}

\def\rv{{\bf r}}

\allowdisplaybreaks
\begin{document}

\title{Optomagnonics in Magnetic Solids}

\author{Tianyu Liu}

\affiliation{Optical Science and Technology Center and Department of Physics and
Astronomy, University of Iowa, Iowa City, Iowa 52242, USA}

\author{Xufeng Zhang}
\affiliation{Department of Electrical Engineering, Yale University, New Haven, CT 06520, USA}

\author{Hong X. Tang}
\affiliation{Department of Electrical Engineering, Yale University, New Haven, CT 06520, USA}

\author{Michael E. Flatt\'{e}}
\affiliation{Optical Science and Technology Center and Department of Physics and
Astronomy, University of Iowa, Iowa City, Iowa 52242, USA}

\begin{abstract}
Coherent conversion of photons to magnons, and back, provides a natural mechanism for rapid control of interactions between stationary spins with long coherence times and high-speed  photons.  Despite the large frequency difference between optical photons and magnons,  coherent conversion can be achieved through a three-particle interaction between one magnon and two photons whose frequency difference is resonant with the magnon frequency, as in optomechanics with two photons and a phonon. The large spin density of a transparent ferromagnetic insulator (such as the ferrite yttrium iron garnet) in an optical cavity provides an intrinsic photon-magnon coupling strength that we calculate to exceed reported optomechanical couplings. A large cavity photon number and properly selected cavity detuning produce a predicted effective coupling strength sufficient for observing electromagnetically induced transparency and the Purcell effect, and even to reach the ultra-strong coupling regime.
\end{abstract}
\maketitle
\newpage

Cavity optomechanics, the optical control of mechanical excitations, has formed the framework for demonstrations of  slow light~\cite{safavi-naeini_electromagnetically_2011} and squeezed light~\cite{purdy_strong_2013}, and proposals for quantum memory~\cite{stannigel_optomechanical_2012}. In cavity optomechanics,  radiation pressure couples the photons in  optical or microwave cavities  to the phonons of mechanical resonators. In addition to clarifying the fundamental nature of quantum interactions and noise, such studies can be applied to  systems in which each excitation provides advantages; e.g., in quantum memory the photons serve as  broad-band long distance information carriers and the phonons as  long-time information storage. Spin waves, as elastic waves, are collective excitations and interact with light. Spin waves, however, are more easily decoupled from the environment than elastic waves, and can also be efficiently manipulated magnetically (in addition to electrically). These advantages suggest a new field,  spin optodynamics, or ``optomagnonics,'' in which optical or microwave fields are paired with these collective spin excitations, whose quanta are known as magnons. Magnons have been shown to efficiently replace radio-frequency (rf) phonons in microwave cavities, in which strong and ultrastrong couplings of magnons and microwave photons have been achieved via the interaction between magnons and the oscillating magnetic fields of the microwave photons~\cite{huebl_high_2013,zhang_strongly_2014,tabuchi_hybridizing_2014,haigh_dispersive_2015,bai_spin_2015}. Recent realizations of weak optical whispering gallery mode coupling to magnetostatic spin waves in a yttrium iron garnet (YIG) sphere are perhaps the first examples of cavity optomagnonics~\cite{haigh_magneto-optical_2015,zhang_optomagnonic_2015,osada_cavity_2015,bourhill_ultra-high_2015}.

\begin{figure}
\includegraphics[width=0.9\linewidth]{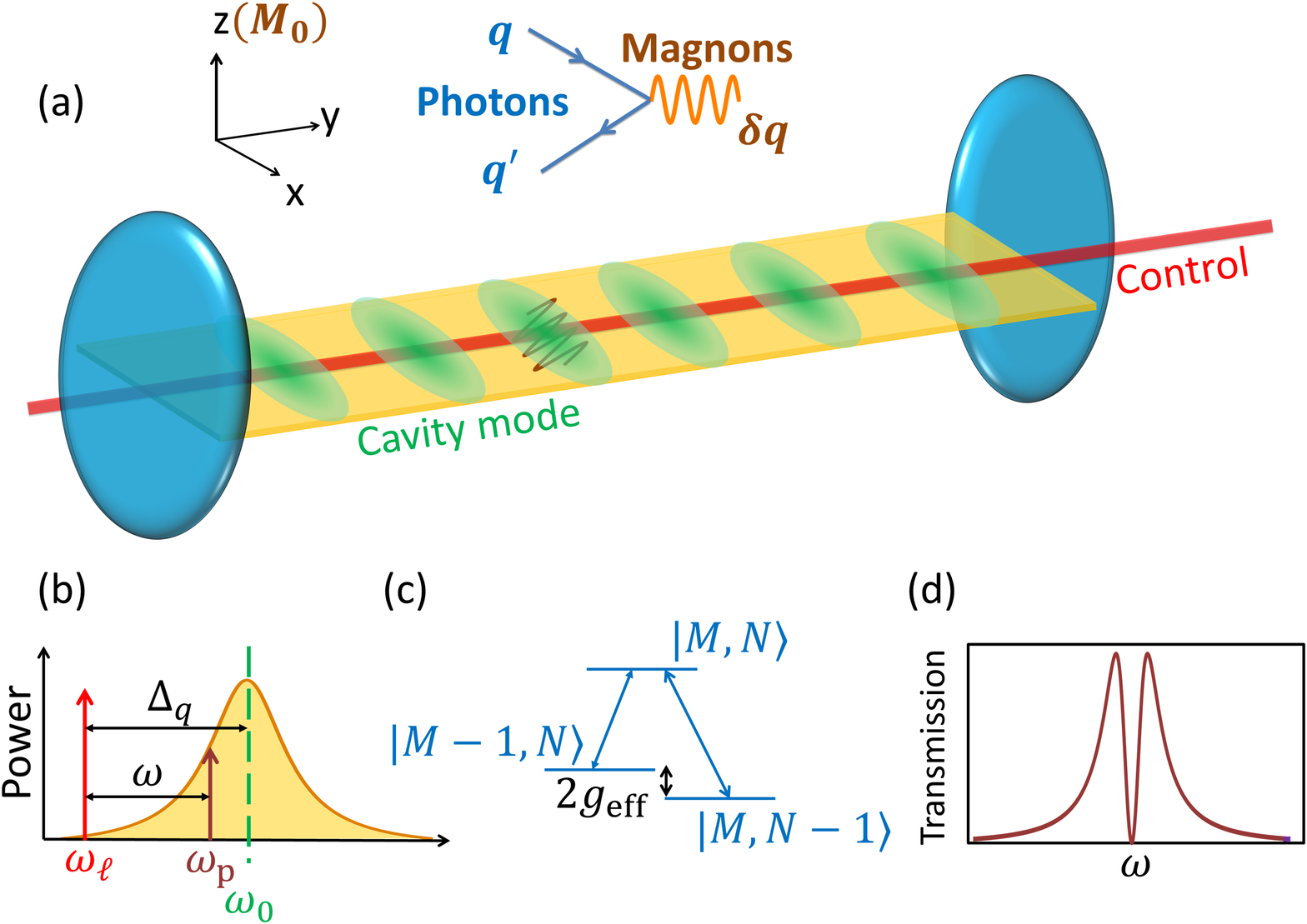}

\caption{(a) Schematic illustration of cavity optomagnonics. The yellow slab sandwiched by two mirrors is the cavity for magnons and for optical photons. (b) $\omega_{\ell}$, $\omega_\text{p}$, and $\omega_0$ denote the frequencies of the control light, the probe light, and the optical cavity mode, respectively. The yellow wave package shows the linewidth of the probe light. (c) $g^{\text{eff}}$ splits two polariton states with different control lights and probe light photon numbers. (d) Due to the three-particle interaction between magnons and photons, the transmission spectra can be tuned by the control light to produce a transparency window around the frequency of the magnon modes.}

\label{Fig1}
\end{figure}

Here, we describe a theoretical framework for optomagnonics, which takes place through the magnon-photon interaction in an optical cavity containing a magnetic slab, as shown in Fig.~\ref{Fig1}. As in cavity optomechanics~\cite{groblacher_observation_2009,hill_coherent_2012,kolkowitz_coherent_2012,dong_optomechanical_2012}, in which the presence of elastic waves modifies the light transmission, here light transmission is modified by the magnetic media and the presence of magnons.
In a microwave cavity  a magnon couples to the magnetic component of the rf fields, and a microwave photon converts directly into a magnon, or vice versa, through a two-particle interaction. For the optomagnonic configuration a three-particle interaction couples a magnon and the electric component of the optical fields within the optical cavity. From the electron-radiation interaction, we calculate the intrinsic magnon-photon coupling strength ($g_0$) in YIG and find that it can be made comparable to or larger than the intrinsic phonon-photon coupling strength in cavity optomechanics. By virtue of detuning and a large photon number, $g_0$ can be enhanced to reach the strong coupling regime where an effective coupling $g^{\text{eff}}$ exceeds either the cavity linewidth or the magnon linewidth, and in these regimes  electromagnetically induced transparency~\cite{safavi-naeini_electromagnetically_2011,zhang_strongly_2014,gouraud_demonstration_2015} and a Purcell enhancement~\cite{purcell,su_towards_2008,zhang_strongly_2014,li_coherent_2015} can be achieved. We find even the ultra-strong regime in which $g^{\text{eff}}$ exceeds both is feasible. These developments in optomagnonics may also assist the low-dissipation propagation of magnons in spintronic devices. For example, an optomagnonic arrangement may form the basis for a high-efficiency, low-dissipation hybrid spintronic interconnect that transmits spin information in optical form. Developments in understanding coherent conversion between magnons and photons may therefore assist  in connecting spintronic devices to a network for quantum communication. Furthermore, the nonreciprocal nature of the magnetic system allows for an isolating, diodelike character of the switching from one mode to another; that is, a first mode can be switched into a second, whereas the second mode does not switch into the first.

\textit{Intrinsic photon-magnon interactions.} Photons interact
with magnons through linear and quadratic magneto-optical coupling,

\begin{eqnarray}
H_{I} & = & \frac{1}{8}\int dV\epsilon_{0}\left\langle \sum_{\alpha,\beta}\epsilon_{r}^{\alpha\beta}(\Mv)E_{1}^{\alpha*}E_{2}^{\beta}+h.c.\right\rangle _{time}\,,
\end{eqnarray}
where $\epsilon_{r}^{\alpha\beta}(\Mv)$ ($\alpha,\beta=x,y,z$) is
the relative dielectric tensor as a function of the magnetization (which includes the effect of magnons).
The time average ensures energy conservation. The subscripts ``1''
and ``2'' denote two light beams that interact with the magnetization,
\begin{eqnarray}
E_{i}^{\alpha} & = & iE_{0i}\sum_{q_{i},m}e_{i,m}^{\alpha}(\xi,\zeta)a_{i,m}^{\dagger}e^{-iq_{i}\eta+i\omega_{i}t}\,,
\end{eqnarray}
where $i=1,2$, $E_{0i}=({2\hbar\omega_{i}}/{\epsilon_{0}n_0^{2}V})^{1/2}$
with $n_0$ the refractive index of YIG, and $\eta$, $\xi$, and $\zeta$ are the coordinates along the length, width, and thickness, respectively. We consider
$\omega_{1}\approx\omega_{2}=(2\pi c_{0})/(n_0\lambda_{0})$,
with $c_{0}$ the speed of light in vacuum, as the excited magnons
have much less energy than the photons. $e_{i,m}^{\alpha}(\xi,\zeta)$
are the normalized field functions for different optical cavity modes,
and $a_{i,m}^{\dagger}$ is the creation operator for cavity mode
$m$. Here, $m$ is a simplified notation for different modes, including distinguishing transverse electric (TE) and magnetic (TM). Although each field is written as a propagating wave, the summation with its Hermitian conjugate yields the appropriate cavity standing wave.  YIG is almost transparent, so we consider only the Hermitian
part of the dielectric tensor. For crystals with cubic symmetry and
assuming the saturation magnetization is along the {[}001{]} direction,
we have, up to linear order in $M_{x}$ and $M_{y}$,
\begin{eqnarray}
\epsilon_{r}(\Mv) & = & \begin{pmatrix}0 & 0 & \epsilon_{r}^{xz}\\
0 & 0 & \epsilon_{r}^{yz}\\
\epsilon_{r}^{zx} & \epsilon_{r}^{zy} & 0
\end{pmatrix}\,,
\end{eqnarray}
with $\epsilon_{r}^{xz}=-iKM_{y}+2G_{44}M_{x}M_{0}$, $\epsilon_{r}^{yz}=iKM_{x}+2G_{44}M_{y}M_{0}$,
$\epsilon_{r}^{zx}=(\epsilon_{r}^{xz})^{*}$, and $\epsilon_{r}^{zy}=(\epsilon_{r}^{yz})^{*}$,
where $\Mv_{0} \parallel \hat z$ is the saturation magnetization~\cite{wettling_relation_1975,wettling_magnetooptical_1975}. $K$ and $G_{44}$ can be obtained from measurements~\cite{wettling_relation_1975,wettling_magnetooptical_1975} of the magnetic
circular birefringence,
\begin{eqnarray}
\Phi_{\text{MCB}} & = & \frac{\pi M_{0}K}{\lambda_{0}n_0}\quad\textrm{for }\kv\parallel\Mv_{0}\,,
\end{eqnarray}
and the magnetic linear birefringence,
\begin{eqnarray}
\Phi_{\text{MLB}} & = & \frac{2\pi M_{0}^{2}G_{44}}{\lambda_{0}n_0}\quad\textrm{for }\Mv_{0}\parallel[111]\perp\kv\,,
\end{eqnarray}
with $\lambda_{0}$ the wavelength of the incident light. Applying the Holstein-Primakoff
transformation~\cite{holstein_field_1940} to the magnetization, we find
\begin{eqnarray}
M^{+}(\rv) & = & \bigg(\frac{2\hbar\gamma M_{0}}{V}\bigg)^{\frac{1}{2}}\sum_{n,k}b_{k}e^{ik\eta}\phi_{n}(\xi,\zeta)\,,\\
M^{-}(\rv) & = & \bigg(\frac{2\hbar\gamma M_{0}}{V}\bigg)^{\frac{1}{2}}\sum_{n,k}b_{k}^{\dagger}e^{-ik\eta}\phi_{n}(\xi,\zeta)\,,
\end{eqnarray}
where $\gamma$ is the gyromagnetic ratio and $\phi_{n}(\xi,\zeta)$ are
the normalized functions for different magnon modes. We assume the cavity
is subject to pinned magnetic boundary conditions on the edges of its cross
section, that is,
\begin{eqnarray}
\phi_{n}(\xi,\zeta) & = & \cos\left(\frac{n\pi\xi}{2w}\right)\,,
\end{eqnarray}
where $n=1,3,5,...$ and $w$ is the half width of the cavity. As the magnon modes we  consider  have frequencies of several GHz, and the dimension along the thickness direction is  small compared with the magnon wavelength, we regard the magnon wave function as homogeneous along the direction of the slab thickness and so it is independent of $\zeta$. The photon-magnon
interaction then simplifies to
\begin{eqnarray}
H_{I} & = & \sum_{m,m',k_{n}}\left[\hbar g_{mm'n}^{(+)}a_{m}a_{m'}^{\dagger}b_{k_{n}}\delta(q_{m}-q_{m'}+k_{n})\right.\nonumber \\
 &  & \left.+\hbar g_{mm'n}^{(-)}a_{m}a_{m'}^{\dagger}b_{k_{n}}^{\dagger}\delta(q_{m}-q_{m'}-k_{n})\right]\,,
\end{eqnarray}
with
\begin{eqnarray}
g_{mm'n}^{(\pm)} & = & \bigg(\frac{2\hbar\gamma}{M_{0}V}\bigg)^{\frac{1}{2}}\frac{c_{0}}{n_0^{2}}
\nonumber \\
 &  & \times
 [\Phi_{\text{MLB}}G_{44,mm'n}^{(\pm)}\pm\Phi_{\text{MCB}}K_{mm'n}^{(\pm)}]\,,
\end{eqnarray}
where
\begin{eqnarray}
G_{44,mm'n}^{(\pm)} & = & \frac{1}{S}\int_{-w}^w d\xi \int_{-d}^d d\zeta\left(e_{1m,z}^{*}e_{2m',x}\mp ie_{1m,z}^{*}e_{2m',y}\right.\nonumber \\
 &  & \left.+e_{1m,x}^{*}e_{2m',z}\mp ie_{1m,y}^{*}e_{2m',z}\right)\phi_{n}(\xi,\zeta)\label{eq:G44}\\
K_{mm'n}^{(\pm)} & = & \frac{1}{S}\int_{-w}^w d\xi \int_{-d}^d d\zeta\left(e_{1m,z}^{*}e_{2m',x}\mp ie_{1m,z}^{*}e_{2m',y}\right.\nonumber \\
 &  & \left.-e_{1m,x}^{*}e_{2m',z}\pm ie_{1m,y}^{*}e_{2m',z}\right)\phi_{n}(\xi,\zeta)\,,\label{eq:K}
\end{eqnarray}
and $S=4wd$ is the area of the cavity cross section. The anti-Stokes and Stokes
processes have different coupling rates $g_{mm'n}^{(\pm)}$, and $g_{mm'n}^{(+)}=g_{m'mn}^{(-)*}$ due to $G_{44,mm'n}^{(\pm)*}=G_{44,m'mn}^{(\mp)}$
and $K_{mm'n}^{(\pm)*}=-K_{m'mn}^{(\mp)}$. This asymmetry has recently been observed in the coupling between whispering gallery modes and magnon modes~\cite{osada_cavity_2015}. Here, we have explicitly written down the dependence of the intrinsic coupling rate $g_{mm'n}^{(\pm)}$ on cavity and magnon mode numbers. For given mode numbers, we will use $g_0^{(\pm)}$ for simplicity instead of $g_{mm'n}^{(\pm)}$.
 $\Phi_{\text{MCB}}\approx\Phi_{\text{MLB}}=6.1$ rad/cm for YIG~\cite{wettling_magnetooptical_1975,stancil_optical-magnetostatic_1991}, so we obtain $g^{(-)}_0\approx2\pi\times 27$ Hz for the Stokes process (TE$_{00}$$\rightarrow$TM$_{31}+\phi_3$ with $\phi_3$ the $n=3$ magnon mode) of $1.55$-$\unit[]{\mu m}$ incident light. A selection rule restricts transitions to those between different polarizations if $\Phi_{\text{MCB}} = \Phi_{\text{MLB}}$. Changing the ratio of $\Phi_{\text{MCB}}$ to $\Phi_{\text{MLB}}$ allows TE to TE and TM to TM transitions but the resulting rates are still much smaller than the TE to TM transitions. The nonreciprocal behavior of transitions from one mode to another is unique to the optomagnonic system, for time-reversal symmetry is broken. In the strong coupling regime discussed later,  this feature and  electromagnetically induced transparency produces an optical diode, in which the probe light of the TE mode is totally reflected into the TM mode, but is absorbed by the cavity for the other modes (including TM to TE). The nature of the control light interaction allows the direction of this optical diode to be switched with the frequency of the control light:  Red detuning allows  TM$\rightarrow$TE but prevents TE$\rightarrow $TM, whereas  blue detuning allows TE$\rightarrow$TM  but prevents TM$\rightarrow$TE.

 The coupling rate also depends on the mode numbers. Along the thickness direction, the two lowest optical modes, together with the homogeneous magnon mode, yield the largest modal overlap. For the width direction, the coupling rates of the opposite-parity TE to TM transitions of other mode numbers differ from that value by less than $1\%$. The parity requirement ensures the integrand in Eqs. (\ref{eq:G44}) and (\ref{eq:K}) is an even function along the width of the cavity.

 The intrinsic photon-magnon coupling strengths in YIG exceed the reported photon-phonon coupling strength ($2\pi\times 2.7\text{ Hz}$) in a cavity that supports strong optomechanical coupling~\cite{groblacher_observation_2009}, suggesting the potential for cavity optomagnonics. Comparing the interaction Hamiltonian in the electro-optical, optomechanical, and optomagnonic systems using the same effective Hamiltonian, $H_I=(\hbar\varphi/\tau)a^{\dagger}a(b+b^{\dagger})$, with $\varphi$ ($\tau$) the optical phase shift (time) of a single round trip, the interaction strength and quality factor $Q_m$ for the three systems are reported in Table~\ref{tab:Comparison} with uniform parameters where possible and typical values for the electro-optical and optomechanical systems~\cite{param}. Table~\ref{tab:Comparison}  shows that, within the same optical cavity, the optomagnonic coupling is ten times larger than the optomechanical coupling, as is the corresponding quality factor. Although the electro-optical coupling is even stronger, the poor quality factor of the fundamental frequency makes strong coupling in an electro-optical system much more difficult than in the other two systems.
\begin{table}[h]

\caption{Comparison of the coupling rates in electro-optical~\cite{tsang_cavity_2010}, optomechanical~\cite{groblacher_observation_2009}
and optomagnonic systems.\label{tab:Comparison}}

\begin{ruledtabular}
\begin{tabular}{cccc}

 & Electro-optical & Optomechanical & Optomagnonic\\
\hline

$\frac{\varphi}{\omega_{0}\tau}$ & $\frac{n_{0}^{2}r}{2d}\sqrt{\frac{\hbar\omega_{m}}{2C}}$ & $\frac{1}{l}\sqrt{\frac{\hbar}{2m\omega_{m}}}$ & $\frac{K}{2n_{0}^{2}}\sqrt{\frac{\hbar\gamma M_{0}}{V}}$\\

values & $1.2\times10^{-11}$ & $3.1\times10^{-14}$ & $3.1\times10^{-13}$\\
$Q_m$ & $500$ & $10^4$ & $10^5$\\

\end{tabular}
\end{ruledtabular}
\end{table}

Spin optodynamics in cold atoms ($^{87}$Rb with a $D_2$ transition)~\cite{brahms_spin_2010,marti_coherent_2014,fang_condensing_2016} has been proposed, so we also compare
the cold atom $g_0$ following our three-particle definition (differing from the convention in Ref.~[\onlinecite{brahms_spin_2010}]) to  YIG.
The linear magneto-optical coupling in the dielectric tensor ($K\epsilon_{\alpha\beta\tau}M_{\tau}$, where  $\epsilon_{\alpha\beta\tau}$ is the Levi-Civita tensor) has the same form for both YIG and the  vector ac-Stark effect in cold atoms. For cold atoms $K=(d^2 v)/(\epsilon_0 \hbar^2\Delta_{ca}\gamma)$ for $^{87}$Rb with $d$ the electric dipole of the $D_2$ transition~\cite{siddons_absolute_2008}, $v$ the vector shift, $\epsilon_0$ the vacuum permittivity, $\Delta_{ca}$ the detuning between the cavity resonance and the $D_2$  transition frequency, and $\gamma$ the gyromagnetic ratio. Although $K(^{87}\text{Rb})/K(\text{YIG})\sim10^7$, the spin density ratio is so small that $M_0(^{87}\text{Rb})/M_0(\text{YIG})\sim10^{-11}$. As $g_0\propto K\sqrt{M_0}$, $g_0(^{87}\text{Rb})$ is of similar order as $g_0(\text{YIG})$.

\textit{Strong coupling regimes}. In analogy to cavity optomechanics, we
consider control light acting on the cavity as shown in Fig.~1(a) to
enhance the photon-magnon interaction by a factor of $\sqrt{N_{\ell}}$
(with $N_{\ell}$ the number of control light photons with frequency
$\omega_{\ell}$). This can be understood in a frame rotating with $\omega_{\ell}$, where $q$ denotes the cavity resonance mode.
Considering the cavity modes $q$ and $q'$ with specific transverse
mode numbers $m$ and $m'$, we will drop the subscripts of $m$ and
$m'$ for simplicity. The system is then described by
\begin{eqnarray}\label{eq:H}
H & = & \sum_{q}\hbar\Delta_{q}a_{q}^{\dagger}a_{q}+\sum_{n,k}\hbar\omega_{nk}b_{k}^{\dagger}b_{k}\nonumber \\
 &  & +\sum_{n,q,q',k}\left[\hbar g_{qq'n}^{(+)}a_{q}a_{q'}^{\dagger}b_{k}\delta(q-q'+k)\right.\nonumber \\
 &  & \left.+\hbar g_{qq'n}^{(-)}a_{q}a_{q'}^{\dagger}b_{k}^{\dagger}\delta(q-q'-k)\right]\,,
\end{eqnarray}
where $\Delta_{q}\equiv\omega_{\ell}-\omega_{q}$ is the detuning of
a control light at frequency $\omega_{\ell}$ from the cavity resonance
frequency $\omega_{q}\equiv\omega_0$, and $a_{q}$ ($b_{k}$) are the annihilation
operators for the optical cavity modes (magnon of frequency $\omega_{k}$).

We derive the
equations of motion for $b_{k}$ and $a_{q}$ from the Heisenberg equation, and find
\begin{eqnarray}
\dot{b}_{k} & = & -(i\omega_{k}+\frac{\gamma_{m}}{2})b_{k}-i\sum_{n,q}g_{q,q-k,n}^{(-)}a_{q}a_{q-k}^{\dagger}\,,\\
\dot{a}_{q} & = & -(i\Delta_{q}+\frac{\kappa_{q}}{2})a_{q}\nonumber \\
 &  & -i\sum_{n,k'}\left[g_{q+k',q,n}^{(-)}a_{q+k'}b_{k'}^{\dagger}+g_{q-k',q,n}^{(+)}a_{q-k'}b_{k'}\right]\nonumber \\
 &  & -({\kappa_{e,q}}/{2})^{1/2}a_{\text{in},q}(t)-{\kappa'^{1/2}_{q}}a_{i,q}(t)\,,
\end{eqnarray}
with $\gamma_{m}=\omega_{m}/Q_{m}$ the magnon damping rate, $\kappa_{e,q}$
the optical damping rate for cavity mode $q$, $\kappa_{q}'$ the parasitic
optical damping rate  into
all other channels that are undetected (representing a loss of information),
and $\kappa_{q}$ the total optical damping rate of mode $q$ ($\kappa_{q}=\kappa_{e,q}/2+\kappa_q'$).
Sources of $\kappa_{q}'$ include homogeneous broadening due to a large
linewidth of the cavity resonance allowing a direct conversion of
the control laser into the cavity resonance mode, or an inhomogeneous
broadening due to absorption by the mirrors of the cavity.

Introducing fluctuations ($\delta a_{q}$ and $\delta b_{k}$) to
the steady states ($\bar{a}_{q}$ and $\bar{b}_{k}$) of the optical
modes [$a_{q}(t)=\bar{a}_{q}+\delta a_{q}(\omega)e^{-i\omega t}$]
and the magnon mode [$b_{k}(t)=\bar{b}_{k}+\delta b_{k}(\omega)e^{-i\omega t}$],
we solve the linear Heisenberg-Langevin equations for the fluctuations
in the frequency space, and obtain the cavity mode spectra,
\begin{eqnarray}
\delta a_{q}(\omega) & = & \frac{-(\kappa_{e,q}/2)^{1/2}\delta a_{\text{in},q}(\omega)-{\kappa'^{1/2}_{q}}a_{i,q}(\omega)}{i(\Delta_{q}-\omega)+\kappa_{q}/2-\kappa^{(-)}}\label{eq:delta-a}
\end{eqnarray}
and $\delta a_{q}^{\dagger}(\omega)=\left[\delta a_{q}(-\omega)\right]^{*}$.
For the lower sideband of the probe light ($\omega\equiv\omega_{\text{p}}-\omega_{\ell}<0$ with $\omega_{\text{p}}$ the frequency of the probe light), $\delta a_{q}(\omega)$
and $\delta a_{q}^{\dagger}(\omega)$ are resonant when the control
light is red detuned ($\Delta_{q}=\omega$) and blue detuned ($\Delta_{q}=-\omega$),
respectively. The input-output boundary conditions $\delta a_{\text{out},q}(\omega)=\delta a_{\text{in},q}(\omega)+(\kappa_{e,q}/2)^{1/2}\delta a_{q}(\omega)$
and $\delta a_{\text{out},q}^{\dagger}(\omega)=\delta a_{\text{in},q}^{\dagger}(\omega)+(\kappa_{e,q}/2)^{1/2}\delta a_{q}^{\dagger}(\omega)$
yield the reflection amplitudes
\begin{eqnarray}\label{eq:t-}
r_{q}^{(-)}(\omega) & = & 1-\frac{\kappa_{e,q}/2}{i(\Delta_{q}-\omega)+\kappa_{q}/2-\kappa^{(-)}}
\end{eqnarray}
for red detuning, and
\begin{eqnarray}\label{eq:t+}
r_{q}^{(+)}(\omega) & = & 1-\frac{\kappa_{e,q}/2}{-i(\Delta_{q}+\omega)+\kappa_{q}/2-\kappa^{(+)}}
\end{eqnarray}
for blue detuning, where
\begin{eqnarray}
&&\kappa^{(-)}(\omega)\nonumber\\
 =&& \sum_n\left[-\frac{\left|g_{q,q-k,n}^{(-)}\right|^{2}N_{\ell,q-k}}{i(\omega_{k}-\omega)+\gamma_{m}/2}+\frac{\left|g_{q+k,q,n}^{(-)}\right|^{2}N_{\ell,q+k}}{-i(\omega_{k}+\omega)+\gamma_{m}/2}\right]\,\,\,\label{eq:kappa+}
\end{eqnarray}
and $\kappa^{(+)}(\omega)=\left[\kappa^{(-)}(-\omega)\right]^{*}$,
with $N_{\ell,q\pm k}=|\bar{a}_{q\pm k}|^{2}$ the photon number of control
light with frequency $\omega_{l,q\pm k}$. The imaginary (real) part
of $\kappa^{(\pm)}$ yields a correction to the resonance frequency
(the line width) of the cavity mode due to the photon-magnon interaction.
The interaction strength is enhanced by $\sqrt{N_{\ell,q\pm k}}$, as
shown in Eq.~(\ref{eq:kappa+}). Thus
\be
g_{q,q-k,n}^{\text{eff}}=g_{q,q-k,n}^{(-)}\sqrt{N_{\ell,q-k}},
\ee
so even though  the intrinsic coupling rate $g_{q,q-k,n}^{(-)}$ is inversely proportional to $\sqrt V$, $g_{q,q-k,n}^{\text{eff}}$  is  proportional to the control light power and is  limited by the maximum allowable photon density.

\begin{figure}
\includegraphics[width=0.9\linewidth]{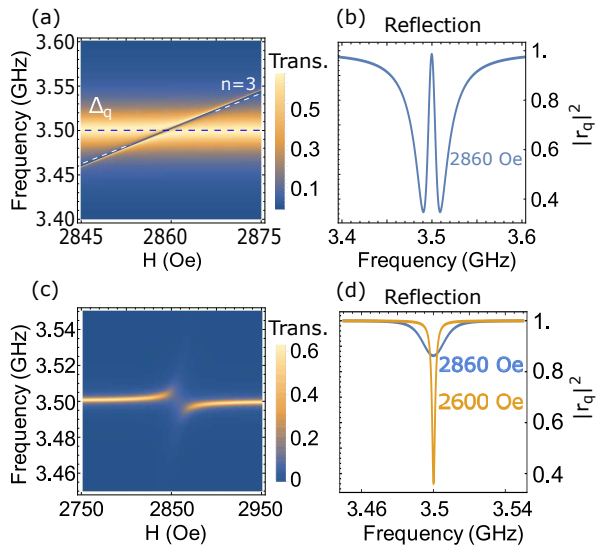}

\caption{Transmission (density plot) and reflection spectra. (a) and (b) Electromagnetically induced transparency (EIT) with $\kappa_q=35$ MHz, $\kappa_{e,q}=14$ MHz, $\gamma_m=0.1$ MHz, $g^{\text{eff}}=10$ MHz, and $\Delta_q=3.5$ GHz. (c) and (d) Purcell enhancement with $\kappa_q=2$ MHz, $\kappa_{e,q}=0.8$ MHz, $\gamma_m=35$ MHz, $g^{\text{eff}}=10$ MHz, and $\Delta_q=3.5$ GHz. The frequency ($\omega$) is the sideband shift of the probe $\omega_{\text{p}}$ from the cavity resonance frequency $\omega_{\ell}$, $\omega=\omega_{\text{p}}-\omega_{\ell}$.
}
\label{Fig2}

\end{figure}

With the control light red detuned from the cavity resonance, we plot the transmission (density plot) and reflection spectra of the anti-Stokes process as shown in Fig. \ref{Fig2}. When $\gamma_m<g^{\text{eff}}<\kappa_q$, one can obtain electromagnetically induced transparency (EIT), as shown in Figs. \ref{Fig2}(a) and 2(b). The asymptotic lines on the density plot denote the resonances with the cavity and with the magnon modes, respectively. The applied magnetic field is swept to tune the magnon mode frequency. When $\omega_k$ is adjusted to be in resonance with the detuned control light, a transparency window is opened in the reflection spectra and its width is determined by $g^{\text{eff}}$.

The EIT properties can be determined by translating Eq. (\ref{eq:H}) to the Tavis-Cummings
model for the fluctuations with the control light red detuned,
\begin{eqnarray}
H & = & \sum_{q}\hbar\Delta_{q}\delta a_{q}^{\dagger}\delta a_{q}+\sum_{n,k}\hbar\omega_{nk}\delta b_{k}^{\dagger}\delta b_{k}\nonumber \\
 &  & +\sum_{n,q,k}\left[\hbar g_{q,q-k,n}^{\text{eff}}\delta a_{q}\delta b_{k}^{\dagger}+h.c.\right]\,,
\end{eqnarray}
which yields two types of polaritons formed by cavity modes dressed
by magnon modes. The energy state of the system can be labeled with the polariton number, $E_{M,N}=\hbar\omega_{+}(N+1/2)+\hbar\omega_{-}(M+1/2)$
with $\omega_{\pm}=(1/2)[(\Delta_{q}+\omega_{k})\pm\sqrt{(\Delta_{q}-\omega_{k})+4|g_{q,q-k,n}^{\text{eff}}|^{2}}]$,
whose energy level diagram is shown in Fig. \ref{Fig1}(c).  $\omega_\ell$ is on resonance
for the $|M,N-1\rangle\rightarrow|M,N\rangle$ transition whereas $\omega_\text{p}$ is
detuned by $\omega$ from the $|M-1,N\rangle\rightarrow|M,N\rangle$
transition. The coexistence of the control and probe light forms
a dark state that makes the probe light less absorbed.

In another regime, for $\kappa_q<g^{\text{eff}}<\gamma_m$, magnons are damped so fast that there are no stable polariton states. On resonance with the magnon modes, the energy of a cavity photon is transferred to a magnon and control photon and then dissipated to the environment. The yellow curve in Fig.~\ref{Fig2}(d) shows the reflection spectrum of the probe light for the off-resonant case, and the blue one for the resonant case. The reflection is strongly reduced when the probe light, interacting with the control light, is resonant with the magnon modes.

\begin{figure}
\includegraphics[width=1\linewidth]{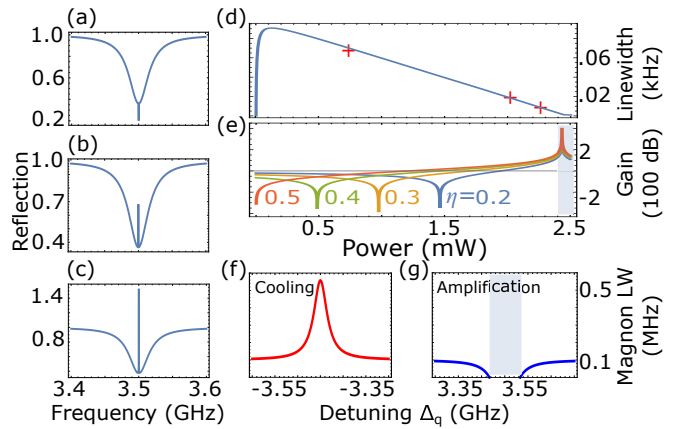}

\caption{(a)-(e) Reflection of the lower side band probe with blue-detuned control at resonance with the magnon mode. (a)-(c) Reflection spectra for different control power [from top to bottom, $0.78$, $2.01$, and $2.26$ mW, which have been labeled with ``+" in (d)]. The frequency ($\omega$) is the sideband shift of the probe $\omega_{\text{p}}$ from the control light $\omega_{\ell}$, $\omega=\omega_{\text{p}}-\omega_{\ell}$. (d) The linewidth of the optomagnonic resonance as a function of the control power. (e) Gain at resonance as a function of the control power with varying $\eta=\kappa_{e,q}/(2\kappa_q)$ for a given $\kappa_q$. The shaded area indicates the instable regime. (f) and (g) The linewidth of the magnon mode as a function of detuning for a given control power ($11$ mW). The red (blue) curve is obtained under the control light of red (blue) detuning. The shaded area indicates the parametric oscillation regime where the linewidth becomes negative. Parameters associated with the plots are $\kappa_q=35$ MHz, $\gamma_m=0.1$ MHz, and $\Delta_q=-3.5$ GHz. $\eta$ is fixed to be 0.2 in the figures other than (e). $\omega_k=3.45$ GHz in (f) and (g).}

\label{Fig3}
\end{figure}

For  blue-detuned control light, the change in the reflection spectra with increasing  control power is shown in Fig.~\ref{Fig3}. The monotonic decrease of the linewidth as  the control light power increases [Fig.~\ref{Fig3}(d)] indicates energy is transferred from the control light to the probe light. The reflection can even exceed one when the gain compensates the energy loss due to the cavity resonance.  Figure~\ref{Fig3}(e) shows that there exists a critical power at which the reflection vanishes. Furthermore, the power is limited according to Eqs.~(\ref{eq:t+}) and (\ref{eq:kappa+}), above which $\kappa^{(+)}=g^{\text{eff}}/(\gamma_m/2)$ cancels the cavity damping rate $\kappa_q$ and the reflection  diverges, leading to a self-oscillation regime. Increasing the  intrinsic optical damping rate moves the system from the undercoupled regime [$\eta=\kappa_{e,q}/(2\kappa_q)<0.25$] to the overcoupled one ($0.25<\eta\leq0.5$), and the critical power decreases so that the reflection becomes divergent more gradually,  yielding more ability to control the effect experimentally~\cite{zhang_cavity_2016}.

To see the direction of power flow in the cavity, we plot the magnon linewidth as a function of detuning for a given control power in Figs. \ref{Fig3}(f) and 3(g), considering the case with $\omega>0$ only, which corresponds to Stokes (anti-Stokes) process for blue (red) detuning. Figure~\ref{Fig3} shows that, for blue detuning, the linewidth decreases as the detuning approaches the resonance with the magnon mode, which indicates that power flows from the control light to magnons, and that even parametric pumping of magnons can be achieved within the shaded range of the detuning. In contrast, for red detuning the linewidth increases as the detuning approaches the resonance with the magnon mode, which indicates that magnons can decay into the optical modes and thus the spin system can be cooled by detuning the control light. For a given $\gamma_m$ and $\kappa_q$, increasing the control power (and thus $g^{\text{eff}}$) will lead to the ultrastrong limit ($g^{\text{eff}}>\gamma_m, \kappa_q$). We found that the corresponding magnon linewidth will not change qualitatively, but its magnitude will increase by orders of magnitude since it is proportional to $(g^{\text{eff}})^2$.

To conclude, we have studied the photon-magnon interaction in an optical cavity made of a magnetic solid. The interaction is intrinsically greater than for optomechanics, and differs in character from the photon-magnon interaction in an microwave cavity. With control light and detuning of the probe light from the cavity resonance, this system can accomplish  coherent conversion between a cavity mode and a magnon mode, or nonreciprocal conversion between two optical modes. As a basis for further studies of quantum dynamics, two classic coherent situations (electromagnetically induced transparency and the Purcell effect) have been simulated.

We note that different aspects of optomagnonic systems have been investigated in a related work done simultaneously in Ref.~\cite{kusminskiy_coupled_2016}. We acknowledge support of the Center for Emergent Materials, a NSF MRSEC under Award No. DMR-1420451 and DARPA MESO.

\bibliography{Ref}

\begin{thebibliography}{32}
\bibitem [{par()}]{param}%
  \BibitemOpen
  \href@noop {} {}\bibinfo {note} {The optical cavity in the three systems is
  taken to be the same, with $l\times w\times
  d=\unit[8]{mm}\times\unit[400]{\mu m}\times\unit[4]{\mu m}$. For the
  electro-optical system, we use the parameters of lithium niobate, with
  $n_0=2.3$, $r=\unit[308]{pm/V}$, $\omega_m=\unit[2\pi\times9]{GHz}$, and
  $C=\unit[1]{pF}$. For the optomechanical system, $m=\unit[145]{ng}$, and
  $\omega_m=\unit[2\pi\times947]{kHz}$. For the optomagnonic system (made of
  YIG), $K=\unit[4.76\times10^{-9}]{m/A}$, $n_0=2.2$, and
  $M_0=\unit[1.39\times10^5]{A/m}$.}\BibitemShut {Stop}%
\end{thebibliography}

\end{document}